%% file: main.tex
\documentclass[a4paper,11pt]{article}
\pdfoutput=1 

\usepackage{jinstpub} 

\usepackage{titlesec}

\title{\boldmath Radiation Hard 3D Silicon Pixel Sensors for use in the ATLAS Detector at the HL-LHC}
\usepackage{subfig}

\author[a]{A.L.~Heggelund,}
\author[b]{S.~Huiberts,}
\author[a]{O.~Dorholt,}
\author[a]{A.L.~Read,}
\author[a]{O.~Rohne,} 
\author[a]{H.~Sandaker,}
\author[b]{M.~Lauritzen,}
\author[b]{B.~Stugu,}
\author[c]{A.~Kok,} 
\author[c]{O.~Koybasi,}
\author[c]{M.~Povoli,}
\author[d]{M.~Bomben,}
\author[e]{J.~Lange\footnote{Now at X-Spectrum GmbH , Luruper Hauptstraße 1, 22547 Hamburg, Germany},}
\author[f]{A.~Rummler}

\affiliation[a]{Department of Physics, University of Oslo, PB 1048 Blindern, 0316 Oslo, Norway}
\affiliation[b]{Department of Physics and Technology, University of Bergen, 5007 Bergen, Norway}
\affiliation[c]{Department of Microsystems and Nanotechnology (MiNaLab), SINTEF Digital, 0373 Oslo, Norway}
\affiliation[d]{Laboratoire de Physique Nucléaire et de Hautes Energies, (LPNHE) Paris, France}
\affiliation[e]{II. Physikalisches Institut, Georg-August-Universität Göttingen, Friedrich-Hund-Platz 1, 37077 Göttingen, Germany}
\affiliation[f]{CERN, Geneva, Switzerland}

\emailAdd{andreas.heggelund@fys.uio.no}

\abstract{The High Luminosity LHC (HL-LHC) upgrade requires the planned Inner Tracker (ITk) of the ATLAS detector to tolerate extremely high radiation doses. Specifically, the innermost parts of the pixel system will have to withstand radiation fluences above $1\times10^{16}$ $n_{eq}cm^{-2}$. Novel 3D silicon pixel sensors offer a superior radiation tolerance compared to conventional planar pixel sensors, and are thus excellent candidates for the innermost parts of the ITk. This paper presents studies of 3D pixel sensors with pixel size $50 \times 50$ $\mu m^2$ mounted on the RD53A prototype readout chip. Following a description of the design and fabrication steps, Test Beam results are presented for unirradiated as well as heavily irradiated sensors. For particles passing at perpendicular incidence, it is shown that  average efficiencies above 96\% are reached for sensors exposed to fluences of $1\times10^{16}$ $n_{eq}cm^{-2}$ when biased to 80 $V$.}


\begin{document}
'
\maketitle
\flushbottom

\input{Sections/Introduction}

\input{Sections/Sensor_Layout_and_stackup}

\input{Sections/SINTEF_processing}

\input{Sections/Testbeams}

\input{Sections/SummaryConclusions}

\acknowledgments

This work was funded by the Research Council of Norway (RCN).

The measurements leading to these results have been performed at the Test Beam Facility at CERN Geneva (Switzerland) and at the Test Beam Facility at DESY Hamburg (Germany), a member of the Helmholtz Association (HGF).

The authors wish to thank fellow testbeam users for constructive discussions and assistance during the common data-taking effort.


\bibliography{Bibliography.bib}{}
\bibliographystyle{unsrt}



\end{document}

%% file: Sections/Introduction.tex
\section{Introduction}
\label{sec:intro}
The current tracker system of the ATLAS detector at CERN's LHC has been performing extremely well, handling high event
rates while withstanding radiation fluences up to a few  
 $10^{15}$ $n_{eq}cm^{-2}$ \cite{ATLASID}.
 It is now approaching the end of its lifetime, and a replacement is needed for operation in an environment of  unprecedented event rates and background radiation expected at the High Luminosity LHC (the HL-LHC).
Significant technological developments are necessary to meet the requirements of the
HL-LHC, where the Inner Tracker (the ITk) is conceived to be composed of silicon microstrip and pixel sensors \cite{StripsTDR,PixelTDR}.
Particularly challenging is the design of pixel  sensors to be placed very close to the interaction region, where fluences in the vicinity of  $2 \times 10^{16}$ $n_{eq}cm^{-2}$ have to be tolerated while maintaining an efficiency of charged particle detection above 96\% for particles at a normal incidence with respect to the beam axis \cite{PixelTDR}.

The 3D sensor 
architecture proposed in 1997 \cite{Parker97} offers an improved radiation hardness compared to conventional planar sensor architectures. Because of the drastically different electrode design, the distance between electrodes can be reduced while still allowing drift of charge carriers in the active region. The reduction in electrode separation reduces the probability of charge trapping, making this sensor design more radiation tolerant.

In practice, 3D sensor production is extremely challenging, and requires significant investments
and development in sensor manufacturing technology. At      
 SINTEF MiNaLab{\footnote{Department of Microsystems and Nanotechnology (MiNaLab), SINTEF Digital,  0373 Oslo, Norway}} several 
productions of 3D test devices have taken place. Results from the first production were reported in 2009 \cite{Sintef09}. 
Furthermore, sensors compatible with the ATLAS IBL
\cite{IBL}, have been proven to perform (when read out by the FEI4 chip \cite{FEI4}) with an efficiency of up to 99.9\% at bias voltages between 5 V and 15 V. \cite{Dorholt18} 

The purpose of this paper is to present results of efficiency studies of a batch of both unirradiated and irradiated sensor test devices from the SINTEF prototyping Run 4 tested in a pion beam at CERN (Section 4.2) and in an electron beam at DESY (Section 4.3). In this batch, sensors with a variety of designs were produced with two different thicknesses of the active material, 100 and 50 {\textmu}m. 
Most sensors on each wafer were designed for mounting on FEI4 front-end chip and only a very small number of sensors adapted to the 'RD53A' chip \cite{RD53A} were included in the wafer design. In Sections 2 and 3 the sensor design and fabrication details will be presented. Beam tests of unirradiated and irradiated sensors mounted on RD53A readout chips are  presented in Section 4.\footnote{Previous tests of sensors mounted on FE-I4 readout chips can be found in Ref. \cite{Dorholt18}.}

%% file: Sections/Sensor_Layout_and_stackup.tex
\section{Sensor and Wafer Layouts}

\begin{figure}[!htb]
\centering
\begin{minipage}{0.7\textwidth}
    \centering
a    \includegraphics[width=1.0\linewidth]{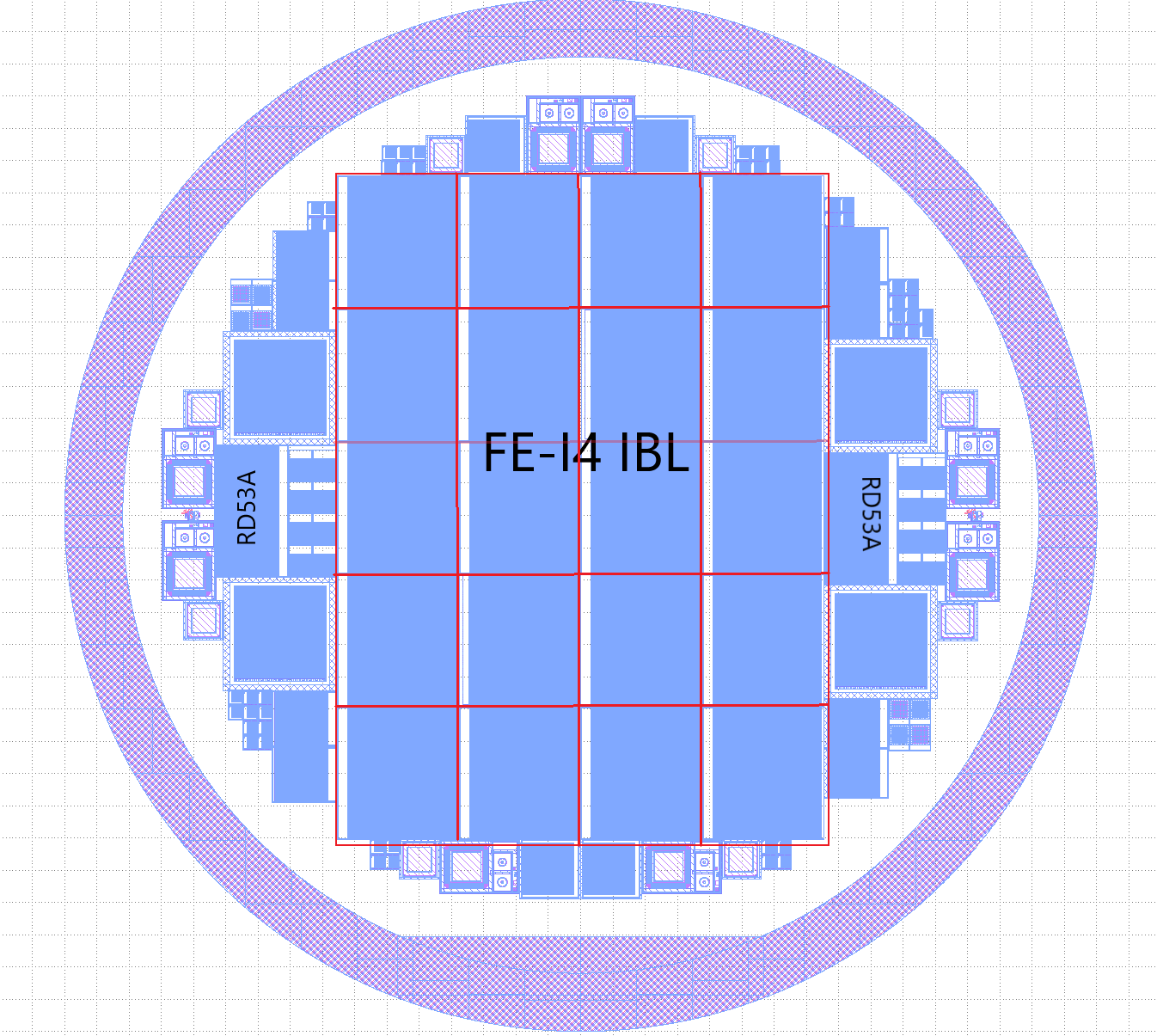}
    \caption{Floor plan of a Run 4 wafer, 20 FE-I4 outlined in red and two RD53A on the sides.}
    \label{fig:floorplan}
\end{minipage}%
\end{figure}

In SINTEF prototype Run 4, eleven 6-inch Si-Si wafers were fully processed. The floorplan of the wafers, which includes two RD53-compatible and 20 FE-I4-compatible sensors, is shown in Figure \ref{fig:floorplan}. This paper presents the test results obtained from tests of RD53A compatible sensors.

\begin{figure}[!htb]
    \centering
    \includegraphics[width=.5\textwidth]{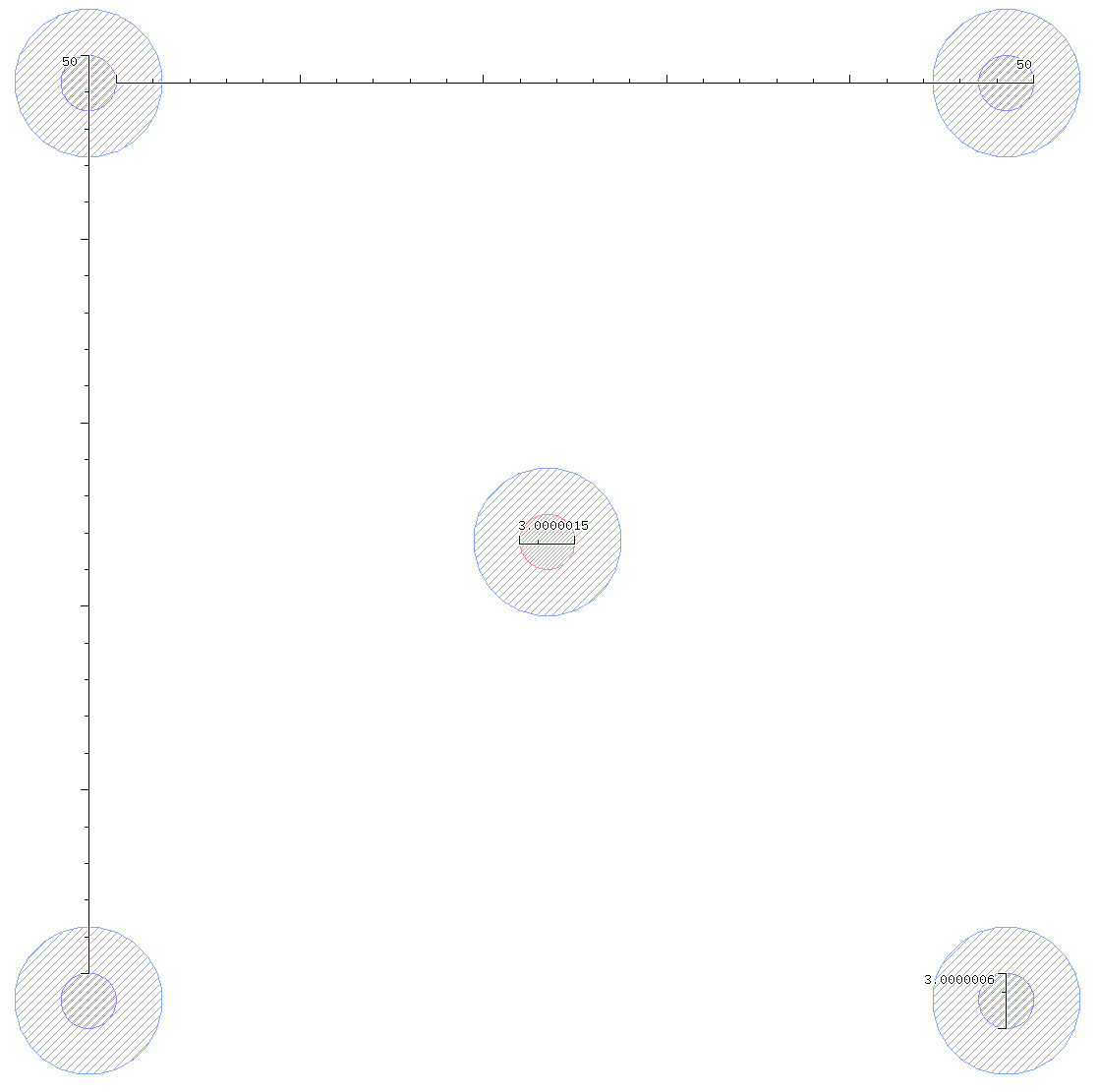}
    
    \caption{Schematic view of a 50x50 $\mu m^2$ pixel cell, with four bias electrodes surrounding the readout-electrode in the middle.}
    \label{fig:schematic_pixelcell}
\end{figure}

The RD53A compatible pixel sensors have a pixel size of $50\times50$ {\textmu}m\textsuperscript{2}. Each pixel consists of one n\textsuperscript{+}-electrode that serves as readout, and four p\textsuperscript{+}-electrodes for biasing, as shown in Figure \ref{fig:schematic_pixelcell}. The readout and biasing electrodes were designed with a diameter of 3 {\textmu}m. Due to the Si-Si wafer, the p\textsuperscript{+}-electrode columns can be contacted on the backside when etched all the way through the active layer, effectively removing the need for high voltage metal on the frontside. 

\begin{figure}[!htb]
    \centering
    \includegraphics[width=.5\textwidth]{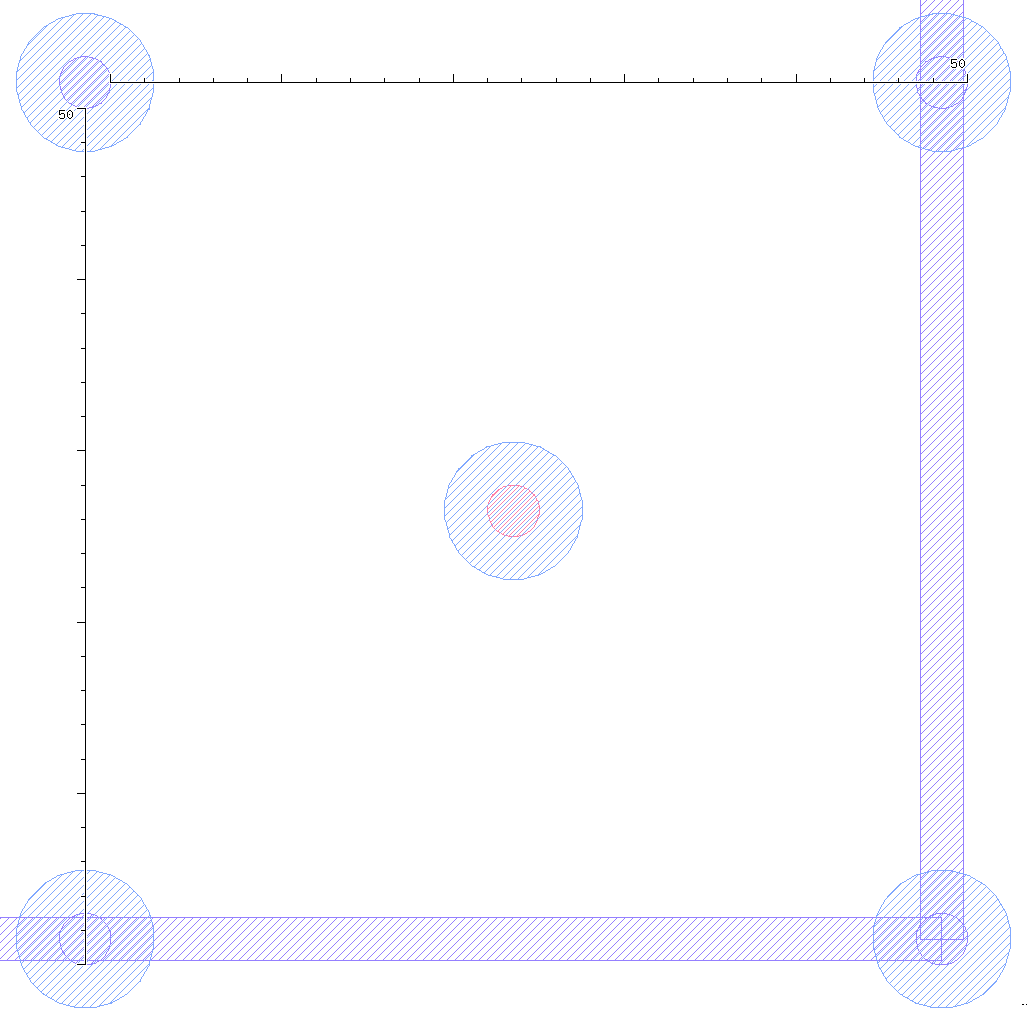}
    
    \caption{Schematic view of a 50x50 $\mu m^2$ pixel cell at a corner of the sensor. The active edge is drawn as the continuous band in the figure.}
    \label{fig:schematic_activeedge}
\end{figure}

As shown in Figure \ref{fig:schematic_activeedge} the sensors are designed with a so-called active edge. This is realized as a continuous electrode trench surrounding and encapsulating the active area of the sensor. This design eliminates the need for guard rings and sensor cutting can be done very close to the active edge, reducing the overall sensor chip size by drastically reducing the dead area.

%% file: Sections/SINTEF_processing.tex
\section{Sensor Fabrication}

The Run 4 fabrication was carried out on 6-inch Si-Si wafers with device layer thicknesses of 100 $\mu m$ and 50 $\mu m$. The device layer is a float zone (FZ), p-type wafer with resistivity of 6000-12000 Ω-cm while the support wafer is a low resistivity p-type wafer. Si-Si wafers, which have no buried oxide between the device wafer and support wafer, offer the capability to contact the p-type electrodes from the backside without additional processing on the support wafer, as compared to Silicon-on-Insulator (SOI) wafers. 

\begin{figure}[!h]

\centering

\subfloat[]{\includegraphics[scale = 0.3]{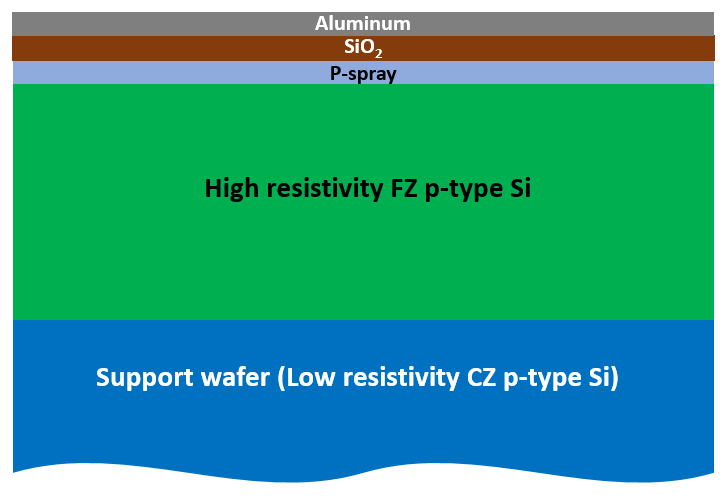}}
\subfloat[]{\includegraphics[scale = 0.3]{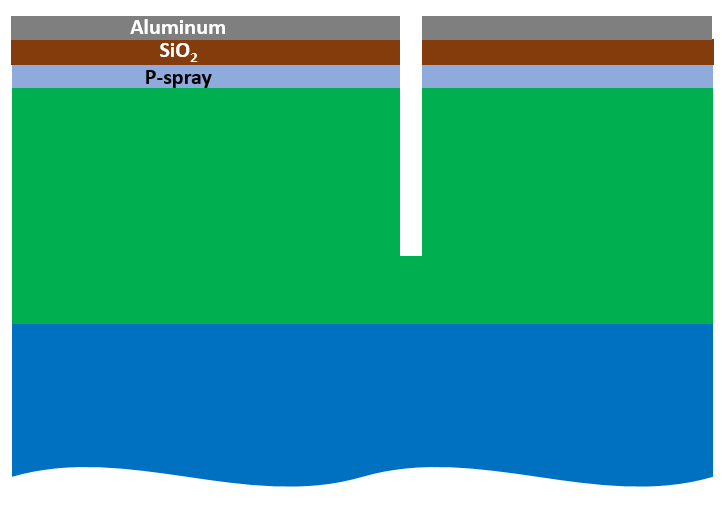}}
\subfloat[]{\includegraphics[scale = 0.3]{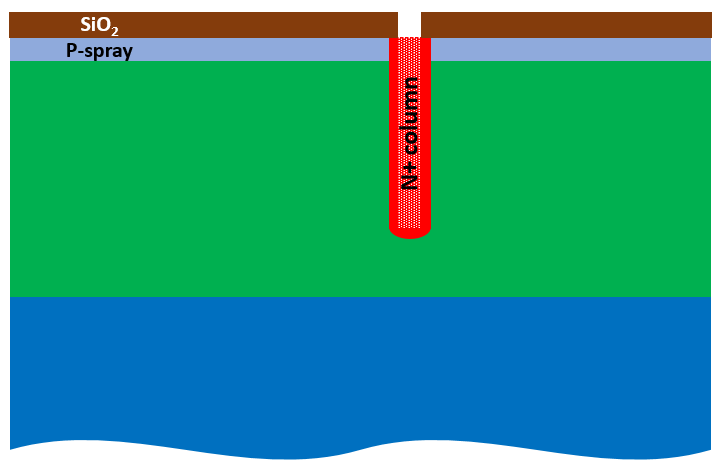}}\hfil
\subfloat[]{\includegraphics[scale = 0.3]{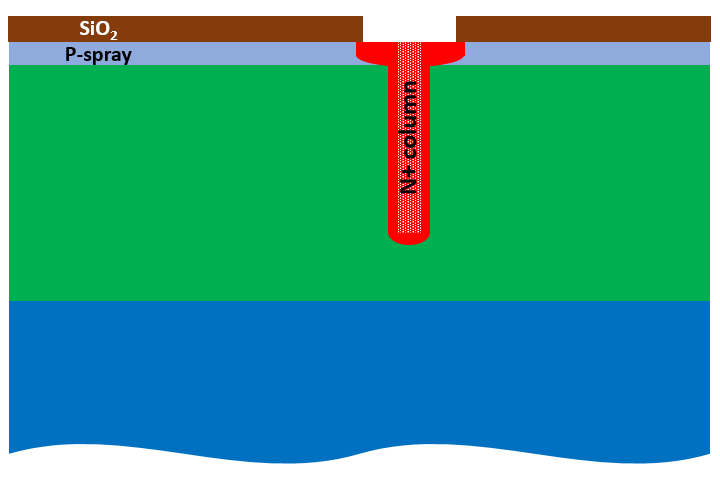}}
\subfloat[]{\includegraphics[scale = 0.3]{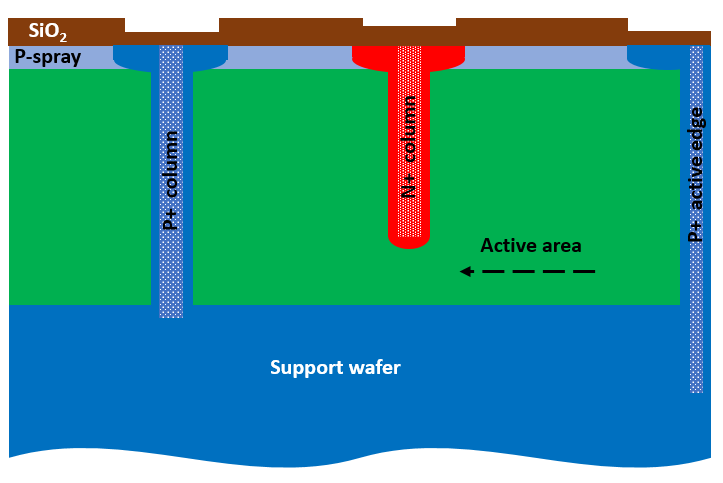}}
\subfloat[]{\includegraphics[scale = 0.3]{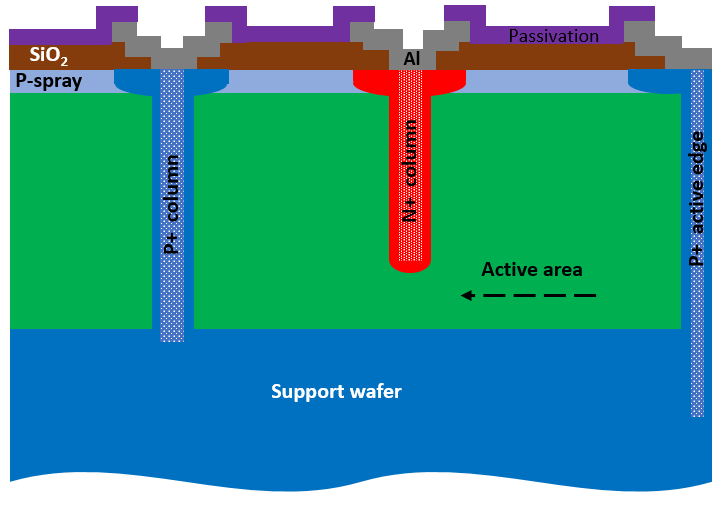}}\hfil
\caption{A brief illustration of the sensor fabrication, \textbf{a)} P-spray implantation, thermal growth of a thick SiO${_2}$ film, deposition of Al mask; \textbf{b)} Patterning of Al mask (n+ column lithography), RIE of SiO${_2}$, DIE of n+ columns; \textbf{c)} Doping of n+ columns, polysilicon deposition, doping of polysilicon, etching of the polysilicon on the surface; \textbf{d)} Patterning of planar n+ regions (n+ surface lithography), n+ doping of the surface; \textbf{e)} Similar processing steps (steps b) to d)) were carried out to fabricate p+ columns and active edge electrodes after another oxidation to cover the n+ regions and deposition of Al mask; \textbf{f)} Standard planar processing (opening of contact holes, metallization, and passivation with PECVD SiO${_2}$+SiN${_x}$).  }\label{fig:FigX}
\end{figure}

A brief summary of the fabrication process is illustrated in Figure \ref{fig:FigX}. The fabrication started with p-spray implantation (low dose boron implantation) to electrically isolate n-type and p-type electrodes. This was followed by some annealing, thermal growth of thick oxide, and sputtering of an aluminum layer used as a hard mask for Deep Reactive Ion Etching (DRIE) of 3D electrodes. Next, the n-type columns were made by patterning the aluminum mask using optical lithography and wet etching, followed by reactive ion etching (RIE) of the oxide and DRIE of silicon. The DRIE was stopped at a reasonably safe distance from the support wafer (35 $\mu m$) to prevent the merging of phosphorous diffused region from the n-type columns and boron diffused region from the support wafer, which would otherwise cause an early breakdown. The aluminum mask was stripped off after the DRIE process was completed, and the columns were doped by gas phase diffusion (POCl3) and filled with phosphorous doped polysilicon to restore the planarity of the wafer so that the subsequent lithographic steps could be carried out. The excess polysilicon deposited on the surfaces of the wafer was removed by RIE. Next, the second lithography process was carried out to etch the oxide around the n-type columns on the wafer surface and dope these surface regions with phosphorous as well to ensure a good electrical contact to n-type columns. Similar processing steps were carried out to fabricate the p-type electrodes and active edge (which are etched all the way down to the support wafer and doped by gas phase boron – BBr3). The sensor fabrication was finished with standard planar processing (contact opening, metallization, and passivation). In total, six wafers with device layer thickness of 100 $\mu m$ and five wafers with device layer thickness of 50 $\mu m$ were completed.

%% file: Sections/Testbeams.tex
\section{Test Beam Measurements}
Sensors equipped with RD53A readout chips were mounted on single-chip cards (SCC) and tested in the H6 pion beam at CERN, as well as in a 4 GeV electron beam at DESY. Sensors with active thicknesses of 50 and 100 microns were tested. Devices under test (DUTs), named D59-1 and E9-1 after their wafer origin \footnote{The letter and number preceding the hyphen reference the wafer from which the sensor was extracted and the subsequent number denotes which number it was given on the wafer.}, were both tested unirradiated. DUTs D61-2 and D62-1 were irradiated to fluences of $n_{eq}=5 \times 10^{15}$ $cm^{-2}$ and $n_{eq} =1 \times 10^{16}$ $cm^{-2}$, respectively, and tested after irradiation.

The RD53A chip is an experimental prototype of the ASICs 
being developed for reading out pixel sensors in the upgraded ATLAS and CMS trackers. Three regions of the chip are equipped with different preamplifiers: 'synchronous', 'linear' and 'differential'. During data-taking, the regions of the sensor equipped with linear and differential preamplifiers were operated and exposed to the beam. The results reported
below are obtained from the region of the sensors that was operated with the differential preamplifier.

\subsection{Test Beam setups}
\begin{figure}[!htb]
    \centering
    \includegraphics[width=.7\textwidth]{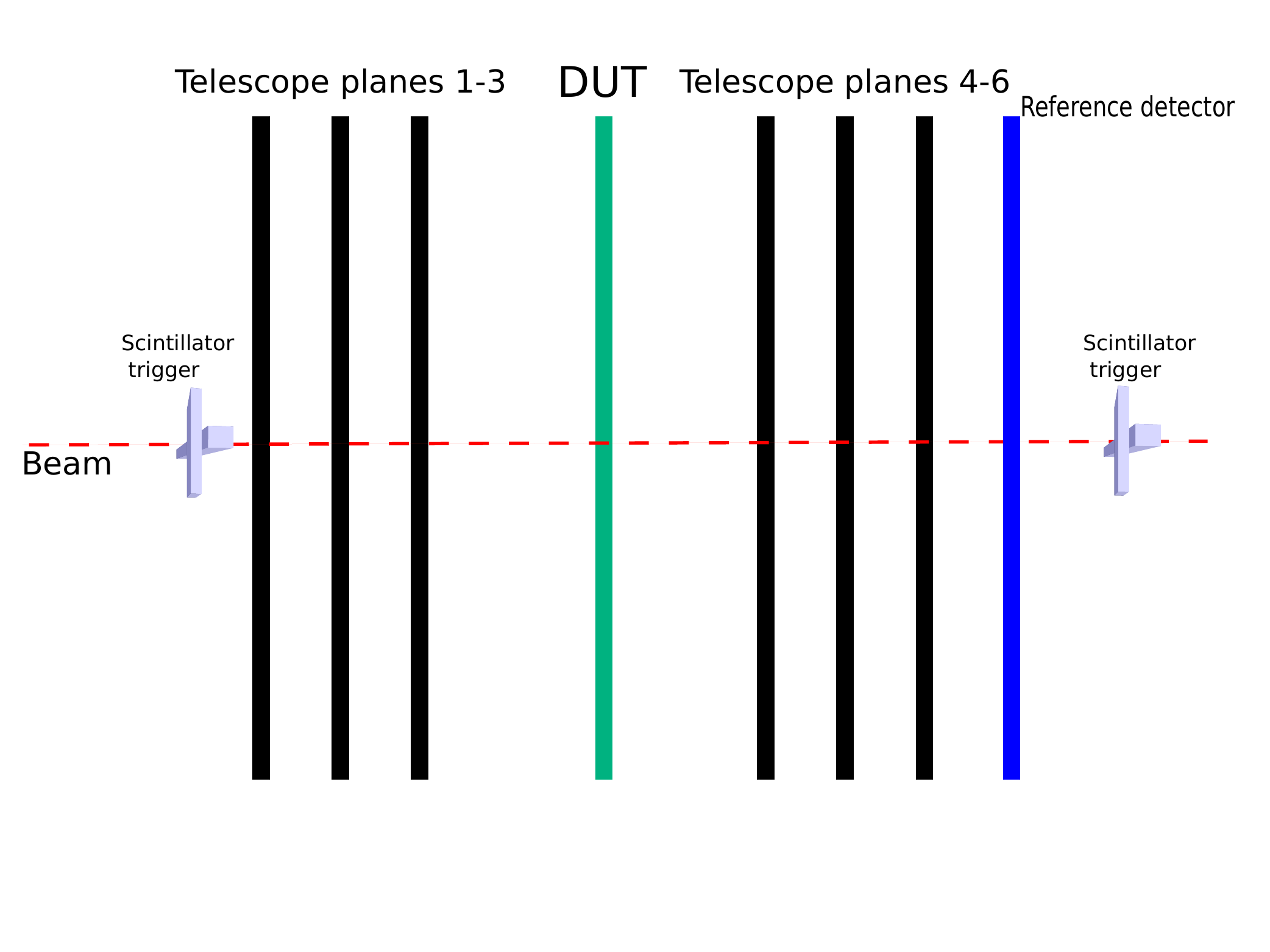}
    \caption{Schematic overview of the testbeam setup. Black lines represent the EUDET telescope planes, the green line represents the DUT, the blue plane shows the placement of the reference detector and scintillation triggers are drawn in at the front and back of the setup. The dashed red line represents the beam.}
    \label{fig:testbeamsetup}
\end{figure}

At CERN and DESY, the devices under test were placed in a setup using the EUDET pixel telescope \cite{EUDET}. The six sensor planes of the telescope are composed of Mimosa26 pixel sensors \cite{MIMOSA} with a pixel pitch of 18.5 {\textmu}m. Each plane consists of 576 x 1152 pixels. The devices under test (DUTs) were placed in the middle of the telescope assembly (three pixel planes upstream and three planes downstream of the DUTs). Triggering was achieved through the use of upstream and downstream scintillators. In addition to the devices under tests, a planar pixel sensor connected to the FE-I4 readout \cite{FEI4} was used as reference. The test beam setup is shown in Figure \ref{fig:testbeamsetup}.

\subsection{Tests in the H6 pion beam at CERN}
\begin{figure}[!htb]
    \centering
    \includegraphics[width=.4\textwidth]{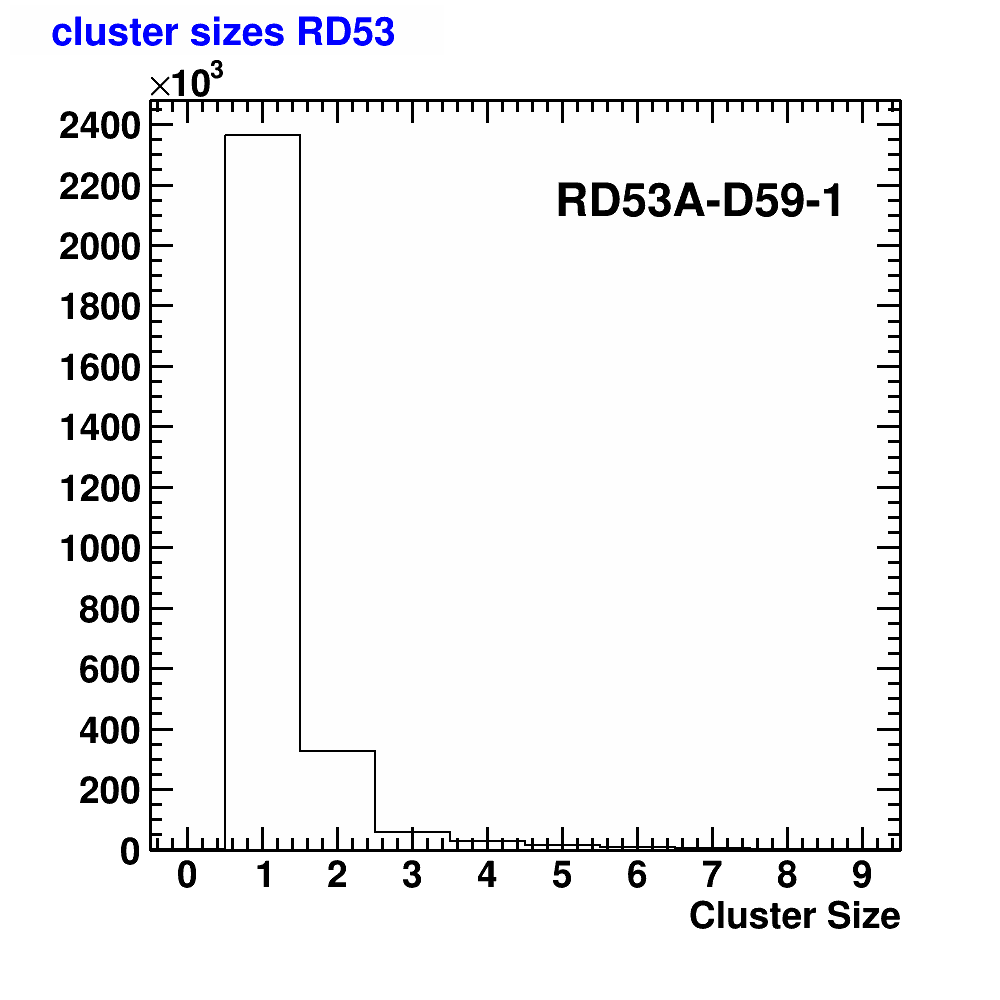}
    \includegraphics[width=.4\textwidth]{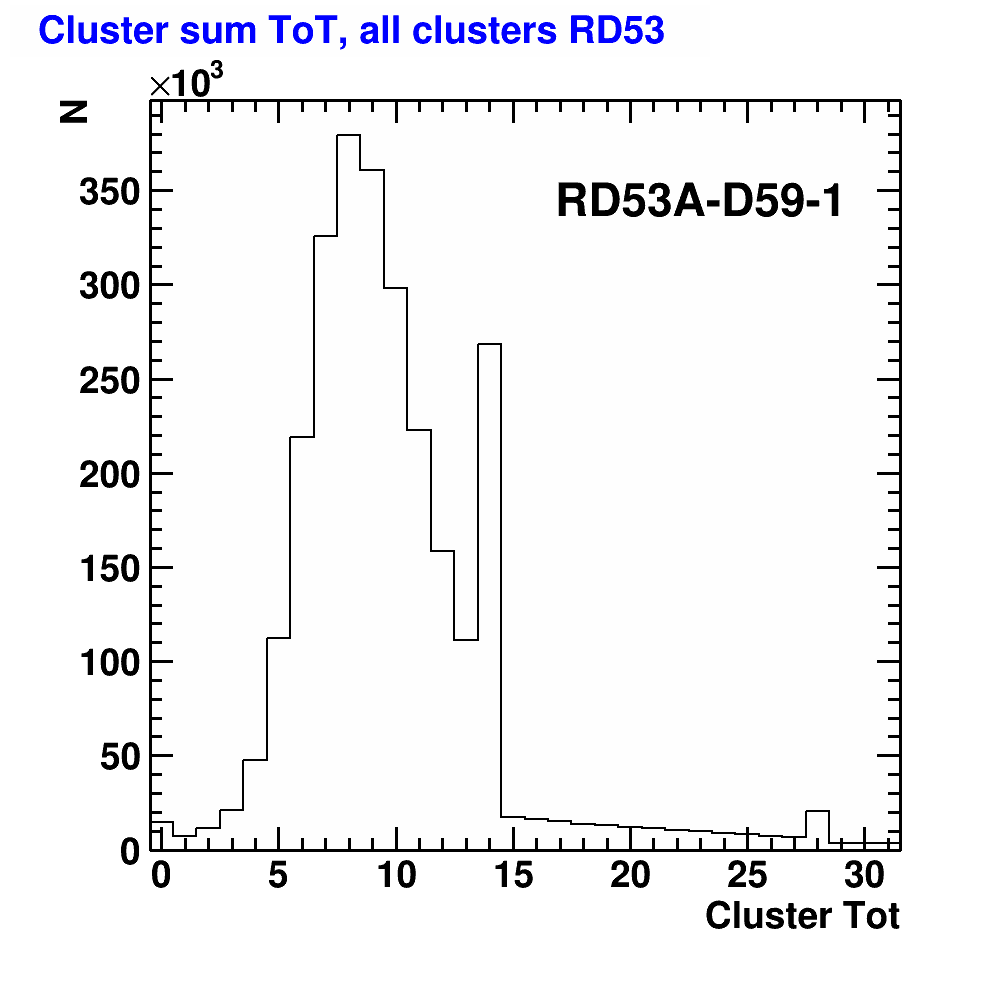}
    \caption{Cluster sizes (left) and summed time over threshold (ToT) for clusters (right) as observed in a pion testbeam, measured at a bias voltage of V$_{bias}$ = 10 V. The rightmost peak in the cluster sum ToT plot, in bin 14, is due to saturation at the highest value for the ToT (14) in a single pixel.}
    \label{fig:TotD59-1_Cern}
\end{figure}

Unirradiated samples were placed in the 180 GeV H6 pion beamline at CERN. In the following, we report on a sensor with 100 {\textmu}m active
thickness, biased at $V_{bias}$ = 10 V, for triggers with a signal present in the reference sensor that could be associated to the reconstructed track. The presence of a signal in the reference sensor is particularly important in this beam, where two particles may arrive in very close succession, with the second particle causing a  new trigger before all tracker data from the previous trigger are cleared. The response of the 100 {\textmu}m thick  sensor, named 'RD53A-D59-1', is shown in 
Figure \ref{fig:TotD59-1_Cern}. Here, the hits from neighbouring pixels are grouped together into clusters and their time over threshold (ToT) are summed up. Most clusters consist of just one hit, but a significant fraction of the events consist of two or more hits (about 15\%).

\begin{figure}[!htb]
\centering
    \begin{minipage}{1.0\textwidth}
    \centering
    \includegraphics[width=.50\textwidth]{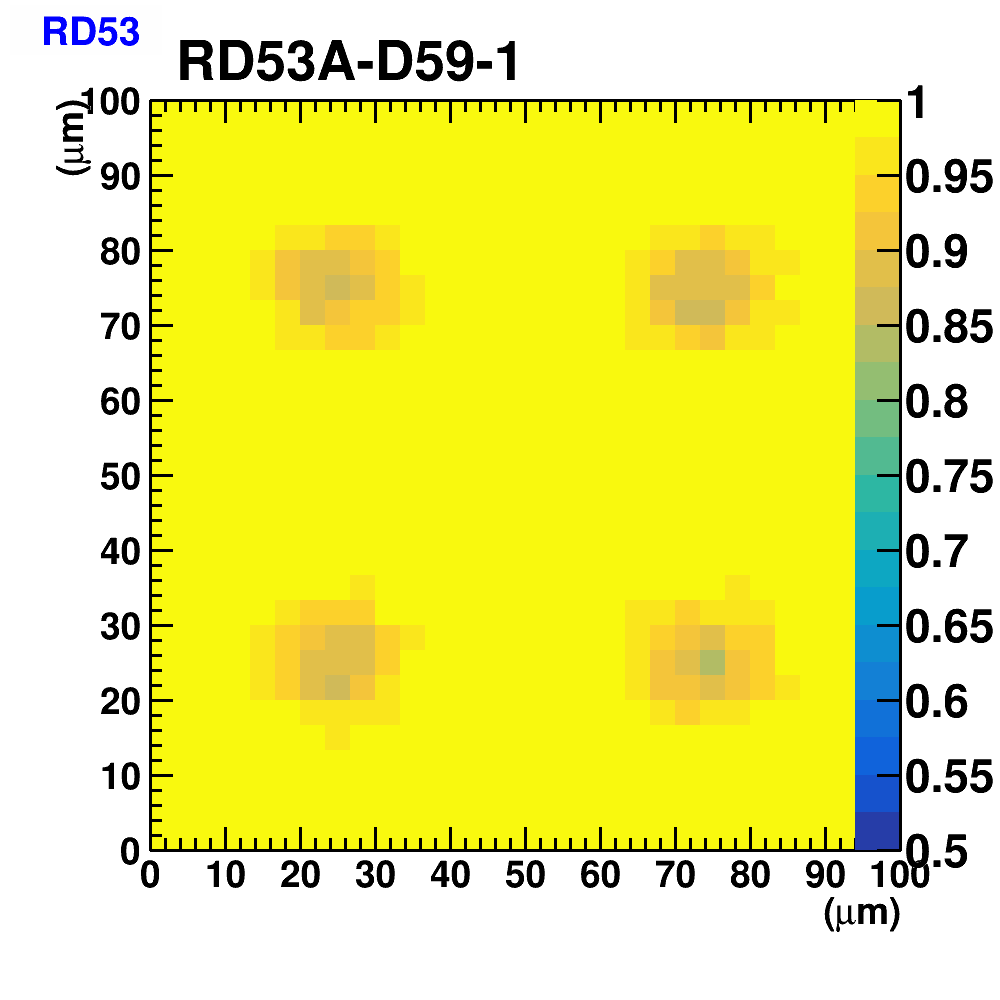}
    \caption{Efficiency map averaged over a region of $100 \times 100 \mu {\rm m}^2$, corresponding to an area covered by four pixels, as measured in the 180GeV pion testbeam at CERN. The size of the  inefficient regions due to the p+ electrodes is consistent with the pointing resolution of the reconstructed tracks.}
    \label{fig:effD59-1_Cern}
\end{minipage}
\end{figure}

The pointing resolution of particle impact points on the DUTs as reconstructed in the CERN test beam was estimated at about 6 {\textmu}m, by the use of distributions of differences between DUT coordinates and track impact predictions. The resolution is sufficient to reveal localized efficiency losses at the locations of the p+ electrodes, as shown in Figure \ref{fig:effD59-1_Cern}. 

The n+ readout electrode is not visible in Figure \ref{fig:effD59-1_Cern} due to the structure of the electrode column. Since this column is not etched all the way through the silicon bulk, charges liberated between the electrode tip and the back side of the sensor are collected. Enough charge is collected that there is no evidence of efficiency loss at the location of the electrode. The average efficiency over a single pixel surface was found to be 98.58\%, with a statistical uncertainty of 0.02\%. Systematic uncertainties include the effects of particle scattering into and out of windows defined for association of hits to reconstructed tracks in the reference sensor as well as in the tested sensor.
Also, in the masking procedures for removing unbonded (dead) and noisy pixels, systematic uncertainties may be introduced. Reasonable variations of the track acceptance-window size, and cuts associated with the definition of noisy pixels, results in total changes in the efficiency estimate well below 0.1\%. Therefore, we quote an estimated efficiency of ($98.6 \pm 0.1 )\%$ for this study.

\subsection{Tests in an electron beam at DESY} 
Both unirradiated and irradiated samples were tested in the 4 GeV electron beam at DESY. As the irradiated samples had to be cooled to avoid thermal runaway from heat dissipation, they were placed in a styrofoam container with dry ice. This method allows for low ambient temperature inside the container but is not stable and did vary between approximately -40 \textdegree{}C and -20 \textdegree{}C. The ambient temperature was monitored visually with a thermometer placed inside the container. The sensor temperature was not recorded during the data-taking, but earlier experiments with the same setup resulted in sensor temperatures between -20 \textdegree{}C and -10 \textdegree{}C.

Care was made to minimize the material at the entry and exit points of the electrons passing through the box and the DUTs. The pointing resolution is dominated by multiple scattering in the material placed in the beam. As the scattering effect is sufficiently high, tracks were reconstructed using the GBL (General Broken Line) algorithm available in the EUtelescope software package. This algorithm is adapted to a situation where most of the material is located in discrete positions. The impact point resolution was found to be 25 to 30 {\textmu}m, depending on the configuration of sensors to be tested, and whether or not the cooling box was present. All the 3D sensors tested at DESY are listed in Table \ref{table:DUTSummary}, together with their corresponding thickness, test parameters, and average efficiencies.

\begin{table}[!htb]
\centering
\caption{List of DUTs measured at DESY testbeam. All DUTS were tuned to a low threshold value between 400-1000 electrons during testbeam.}
\resizebox{\textwidth}{!}{

\begin{tabular}{ccccc}
\multicolumn{5}{c}{{\textbf{Overview of DUTs parameters}}}
\\
\hline
\multicolumn{1}{|c|}{{Name of DUT}} & \multicolumn{1}{c|}{{Thickness {[}$\mu m${]}}} & \multicolumn{1}{c|}{{Threshold{[}$e^-${]}}} & \multicolumn{1}{c|} {{Dose [$n_{eq} cm^{-2}$]}} & \multicolumn{1}{c|}{{Efficiency {[}\%{]}}}
\\
\hline
\multicolumn{1}{|c|}{D59-1} & \multicolumn{1}{c|}{100} & \multicolumn{1}{c|}{400} & \multicolumn{1}{c|}{Unirradiated} & \multicolumn{1}{c|}{99.5} \\
\multicolumn{1}{|c|}{E9-1} & \multicolumn{1}{c|}{50} & \multicolumn{1}{c|}{400} & \multicolumn{1}{c|}{Unirradiated} & \multicolumn{1}{c|}{98.9} \\
\multicolumn{1}{|c|}{D61-2} & \multicolumn{1}{c|}{100} & \multicolumn{1}{c|}{1000} & \multicolumn{1}{c|}{$5\times10^{15}$} & \multicolumn{1}{c|}{above 97\%} \\
\multicolumn{1}{|c|}{D62-1} & \multicolumn{1}{c|}{100} & \multicolumn{1}{c|}{500} & \multicolumn{1}{c|}{$1\times10^{16}$} & \multicolumn{1}{c|}{above 96\%}
\\
\hline
\end{tabular}%
}

\label{table:DUTSummary}
\end{table}

\subsubsection{Studies on non-irradiated samples}
Two non-irradiated sensors,  'E9-1' and 'D59-1', with active thicknesses of 50  and 100 {\textmu}m were studied at different bias voltages between 2.5 V and 20 V. In order to minimize signal loss and stay above noisy low threshold regions, both sensors were tuned to a threshold value of 400 $e^-$. Runs were performed with two different tilt angles with respect to the beam, perpendicular incidence, and a 15\textdegree{} tilt with respect to normal incidence.\ 

\begin{figure}[!htb]
    \centering
    \includegraphics[width=.48\textwidth]{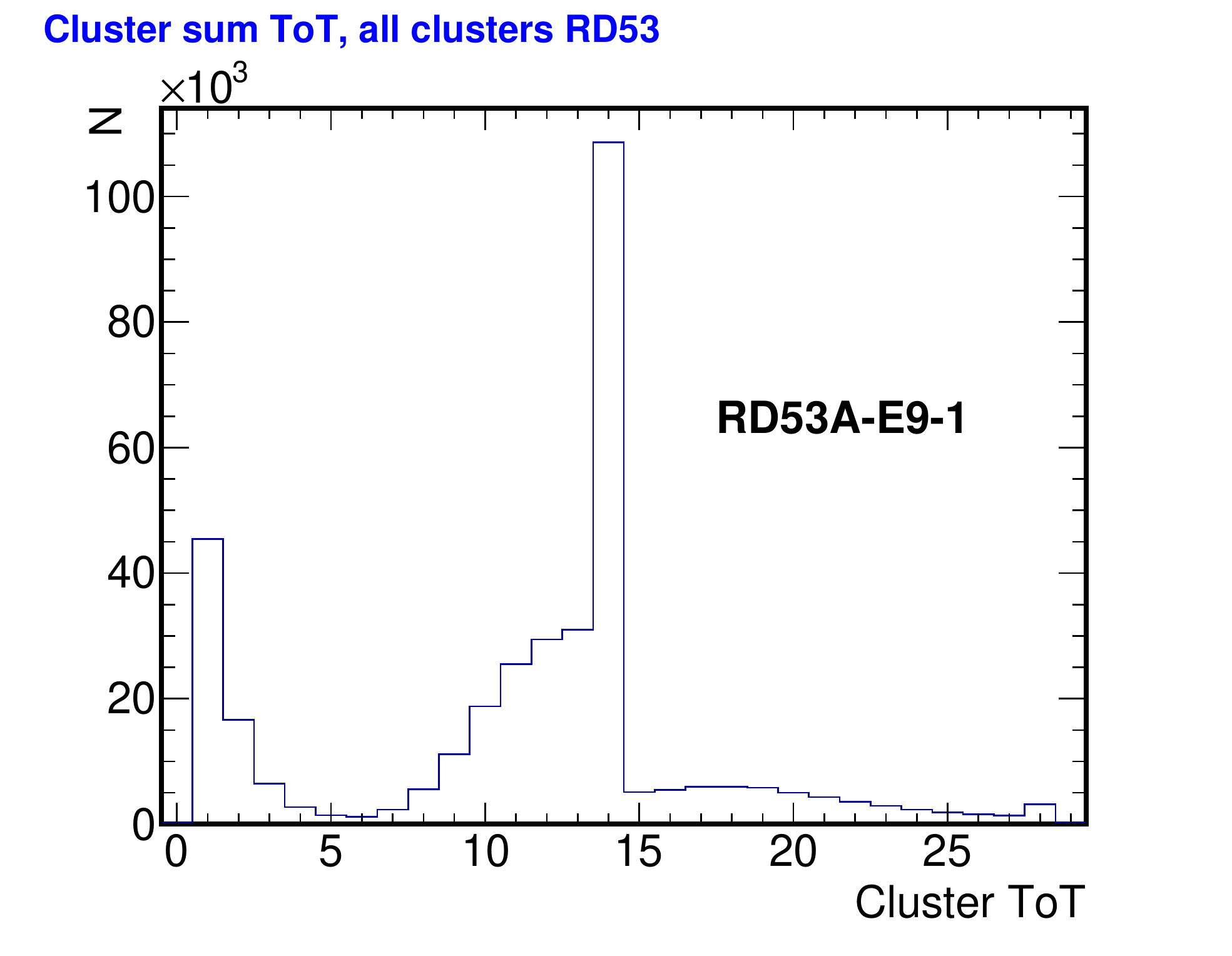}
    \includegraphics[width=.48\textwidth]{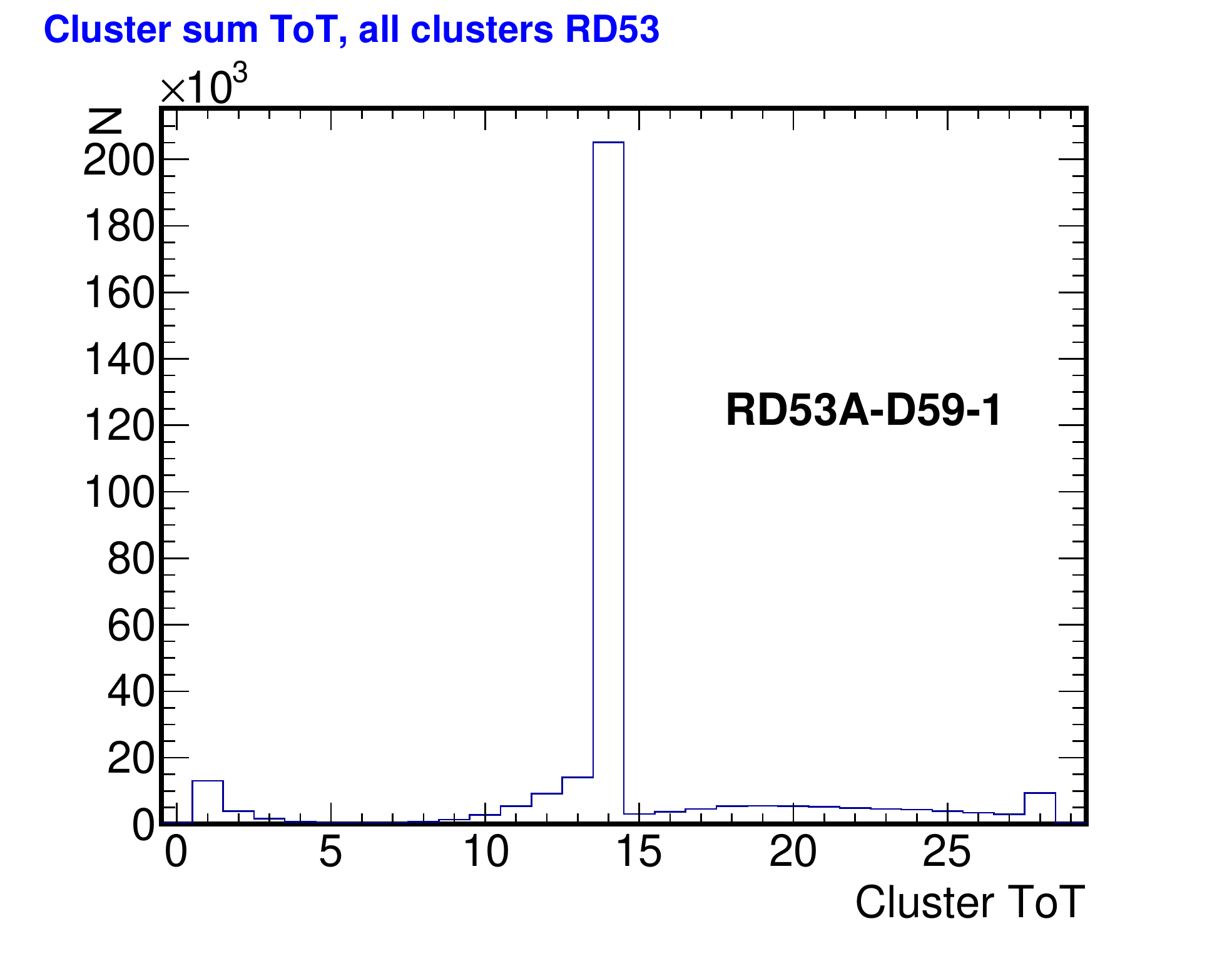}
    \caption{ToT for E9-1 (left) and D59-1 clusters (right) as observed in an electron beam at bias voltage = -20 V. The rightmost peak in the cluster sum ToT plots, in bin 14, is due to saturation at the highest value for the ToT (14) in a single pixel.}
    \label{fig:ToTD59-1_DESY}
\end{figure}

Figure \ref{fig:ToTD59-1_DESY} shows the Time-over-Threshold (ToT) for both the sensors perpendicular to the beam. As seen in these plots, the ToT was not tuned properly before data-taking on the DUTs for the 400 $e^-$ configuration. This caused the ToT distribution to have its centre  above the maximum register value of the chip. Thus, hit saturation leads to an overflow in bin 14, resulting in a high, artificial peak. 

\begin{figure}[!htb]
    \centering
    \includegraphics[width=.48\textwidth]{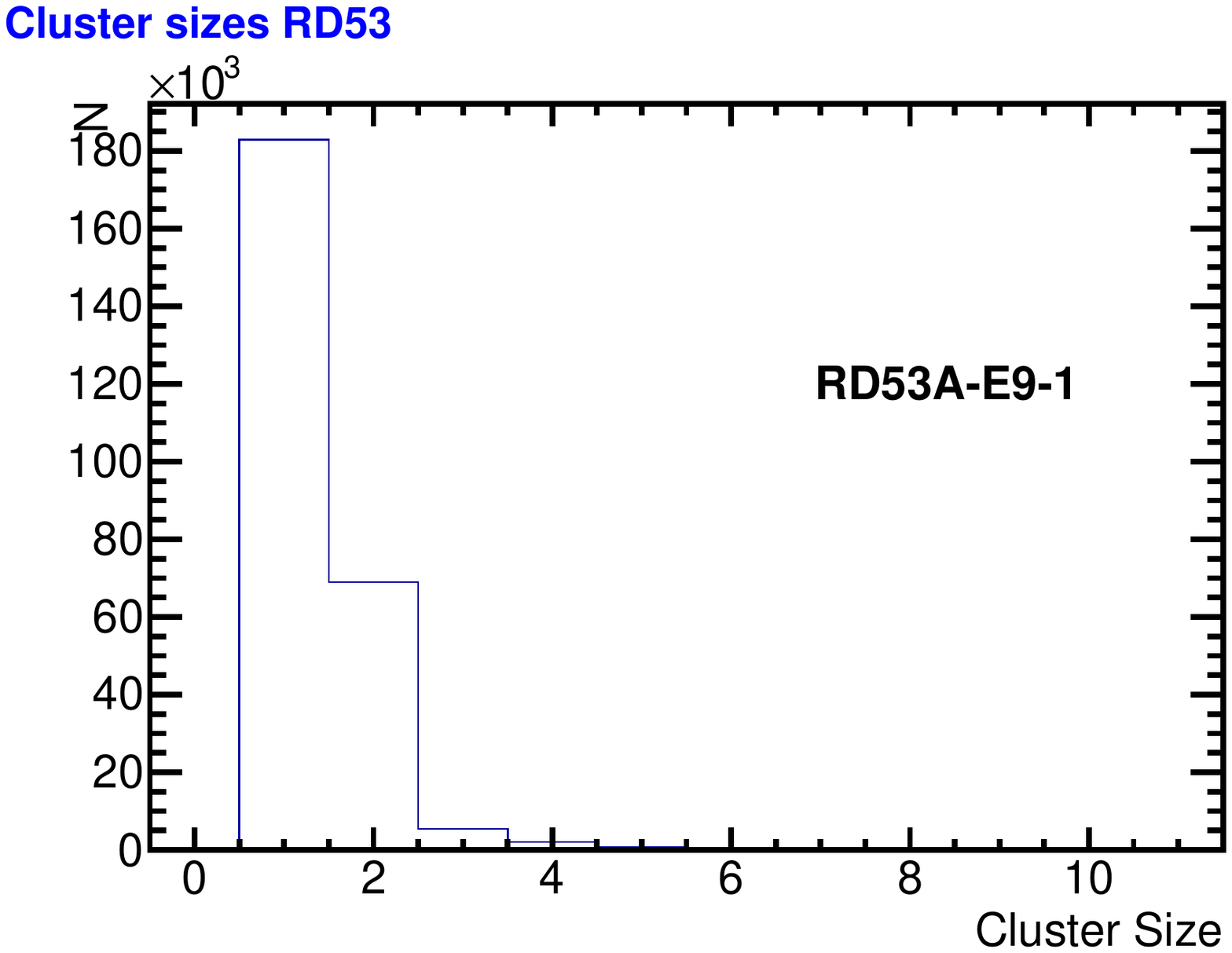}
    \includegraphics[width=.48\textwidth]{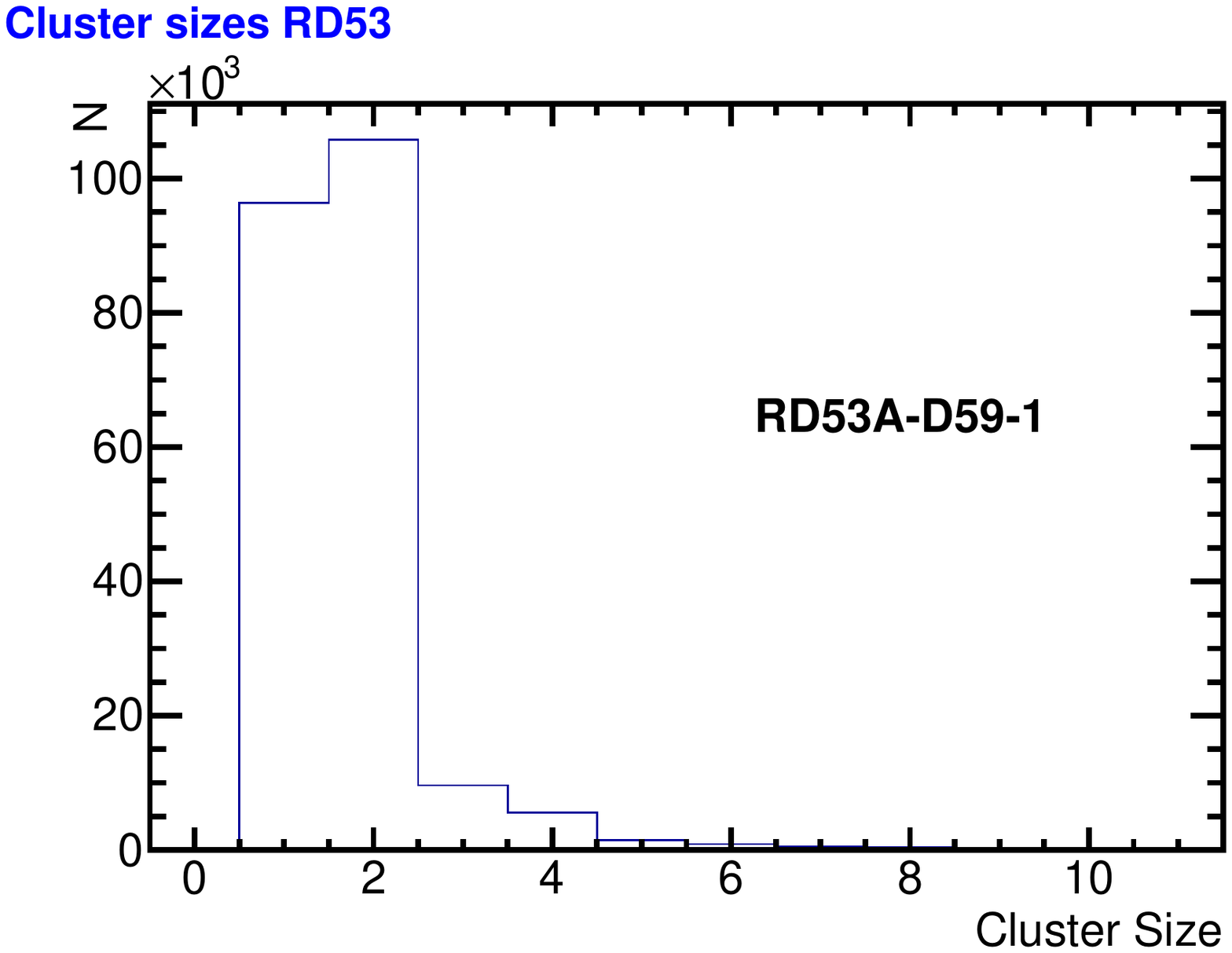}
    \caption{Cluster sizes in number of pixels for E9-1 (left) and D59-1 clusters (right) as observed in an electron testbeam when tilted at 15 degrees with bias voltage = -20 V. }
    \label{fig:TotD59-1_DESY2}
\end{figure}

Figure \ref{fig:TotD59-1_DESY2} shows the cluster sizes for the same sensors but with a 15\textdegree{} tilt with respect to normal incidence. As seen from the two distributions, the thicker sensor (right plot) has a higher relative amount of bigger clusters compared to the thinner sensors (left plot). 

The overflow in bin 14 led to a non-representative ToT distribution and skewed the average ToT value to a lower number. The ToT of the DUTs, however, can still be compared as the effect was the same in both. Post-analysis estimation of the average ToT bin values shows that the mean bin value increased from 12.11 to 12.35 ($\pm 0.01$ for both) for the thinner 50 {\textmu}m 'E9-1' sensor when tilted, while an estimated increase of the mean bin value from 11.328 $\pm$ 0.003 to 15.183 $\pm$ 0.009 was observed for the 100 {\textmu}m thick sensor 'D59-1'. 
This effect can also been seen in the relative cluster size, where the average size of the clusters increased from to 1.246 to 1.368 ($\pm 0.003$ for both) in the thinner sensor, and from 1.314 $\pm$ 0.001 to 1.735 $\pm$ 0.003 in the thicker sensor with and without the tilt, respectively. 

An increase in the ToT values going from a thinner to a thicker sensor as well as an increase in cluster sizes when tilting the sensors are expected. In both cases, the electrons from the beam have a long path-length in the sensitive region of the sensor, thus, cluster sizes and the summed cluster ToT values increases in both sensors with a tilt of 15\textdegree{}.

\begin{figure}[!htb]
    \centering
    \advance\leftskip-1cm
    \includegraphics[width=1\textwidth]{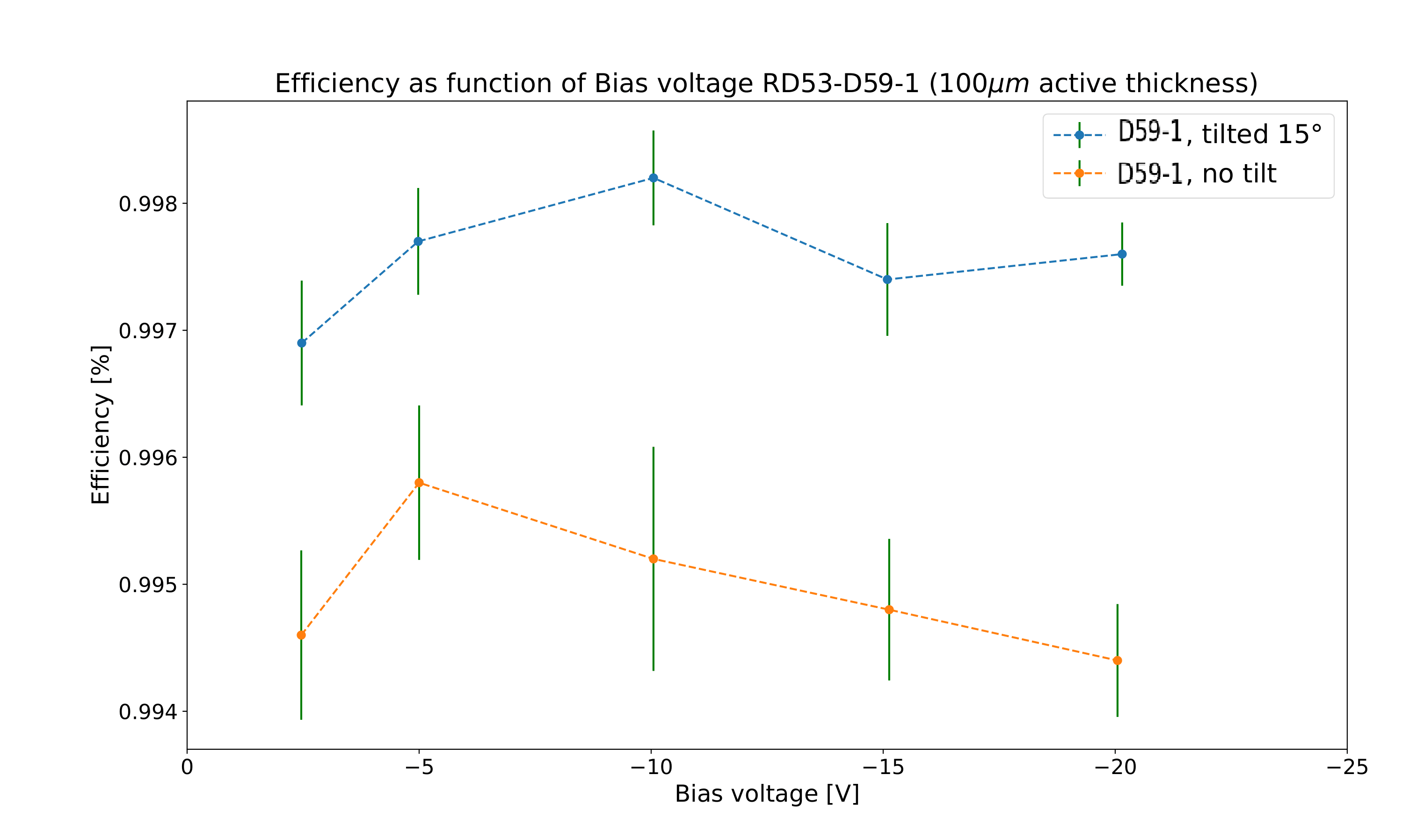}
    \caption{Efficiencies for different bias voltages for a 100 {\textmu}m thick sensor at perpendicular beam incidence (orange), and an incident angle of 15 \textdegree{} (blue).}
    \label{fig:D59-1_eff}
\end{figure}

\begin{figure}[!htb]
    \centering
    \includegraphics[width=1\textwidth]{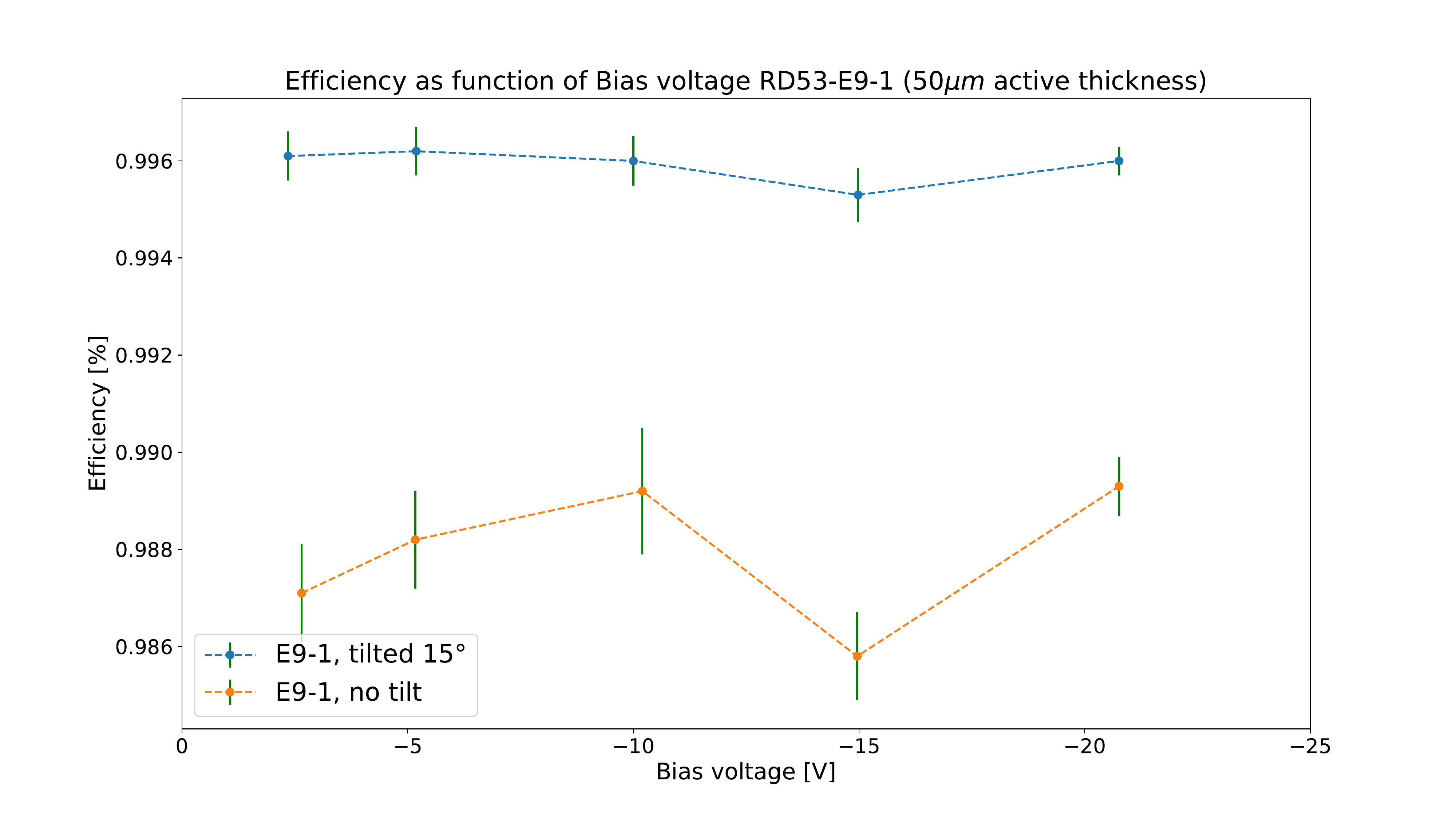}
    \caption{Efficiencies for different bias voltages for a 50 {\textmu}m thick sensor at perpendicular beam incidence (orange), and an incident angle of 15 \textdegree{} (blue).}
    \label{fig:E9-1_eff}
\end{figure}

Figures \ref{fig:D59-1_eff} and \ref{fig:E9-1_eff} summarize efficiency measurements as functions of bias voltage for sensors of 100 {\textmu}m and 50 {\textmu}m, respectively. It is clearly demonstrated that the particle-detection efficiency is higher when the sensor is inclined by 15\textdegree{}. Furthermore, it is seen that the sensors can be operated with full efficiency at a bias as low as 2.5 V. At a bias of 10 V, efficiencies to perpendicular (tilted) tracks are found to be 99.5\% (99.8 \%) for the 100 {\textmu}m sensor, and 98.9 \% (99.6 \%) for the 50 {\textmu}m sensor. The higher efficiency observed in the data collected at DESY is believed to primarily be due to the DUT being tuned to the high amplification described above in combination with the lower charge collection threshold of 400 $e^-$ at DESY compared to 800 $e^-$ at CERN.

\hfil
\subsubsection{Studies of irradiated sensors}
\begin{figure}[!htb]
\centering
\begin{minipage}{.5\textwidth}
  \centering
  \includegraphics[width=1.0\linewidth]{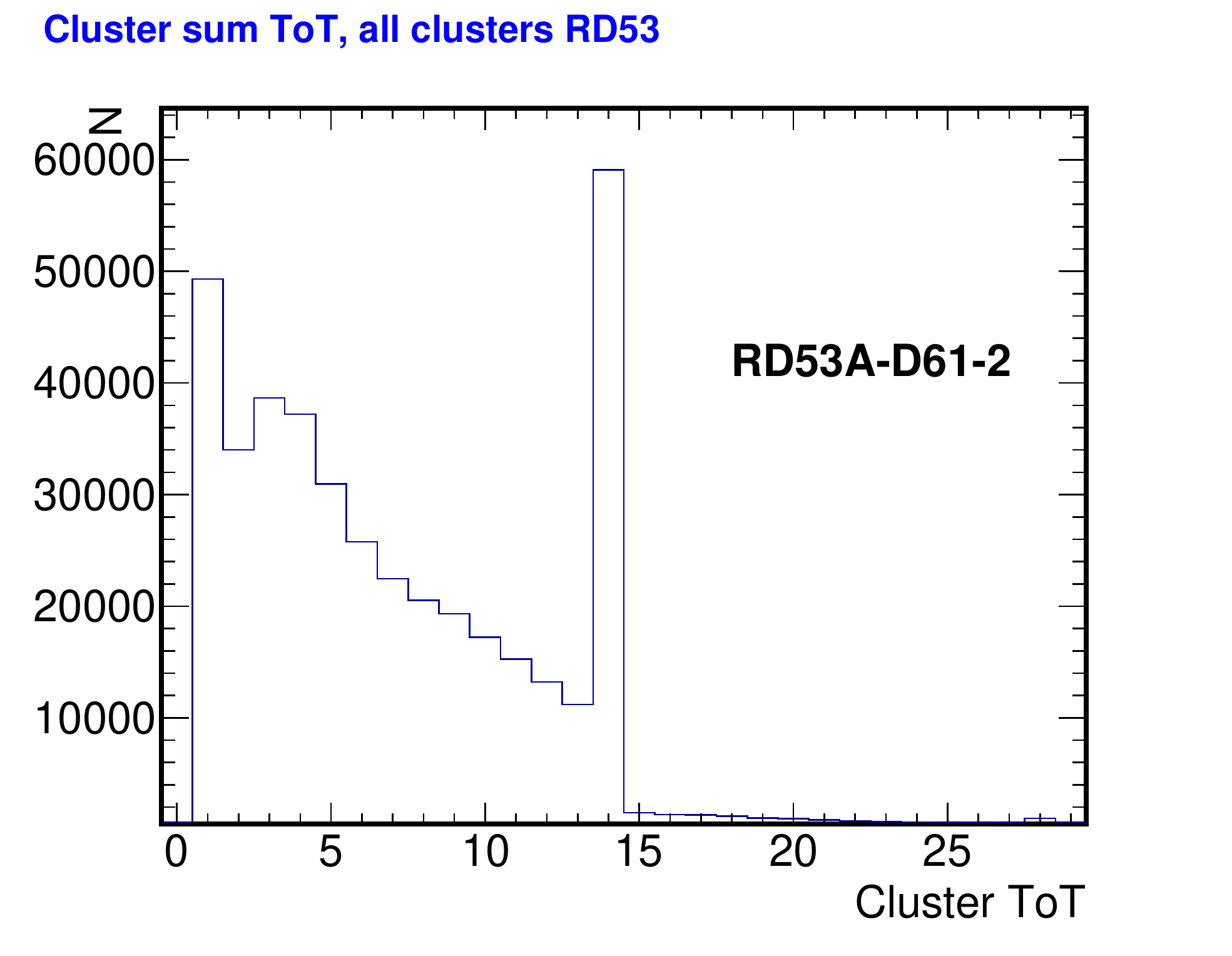}

\end{minipage}%
\begin{minipage}{.5\textwidth}
  \centering
  \includegraphics[width=1.0\linewidth]{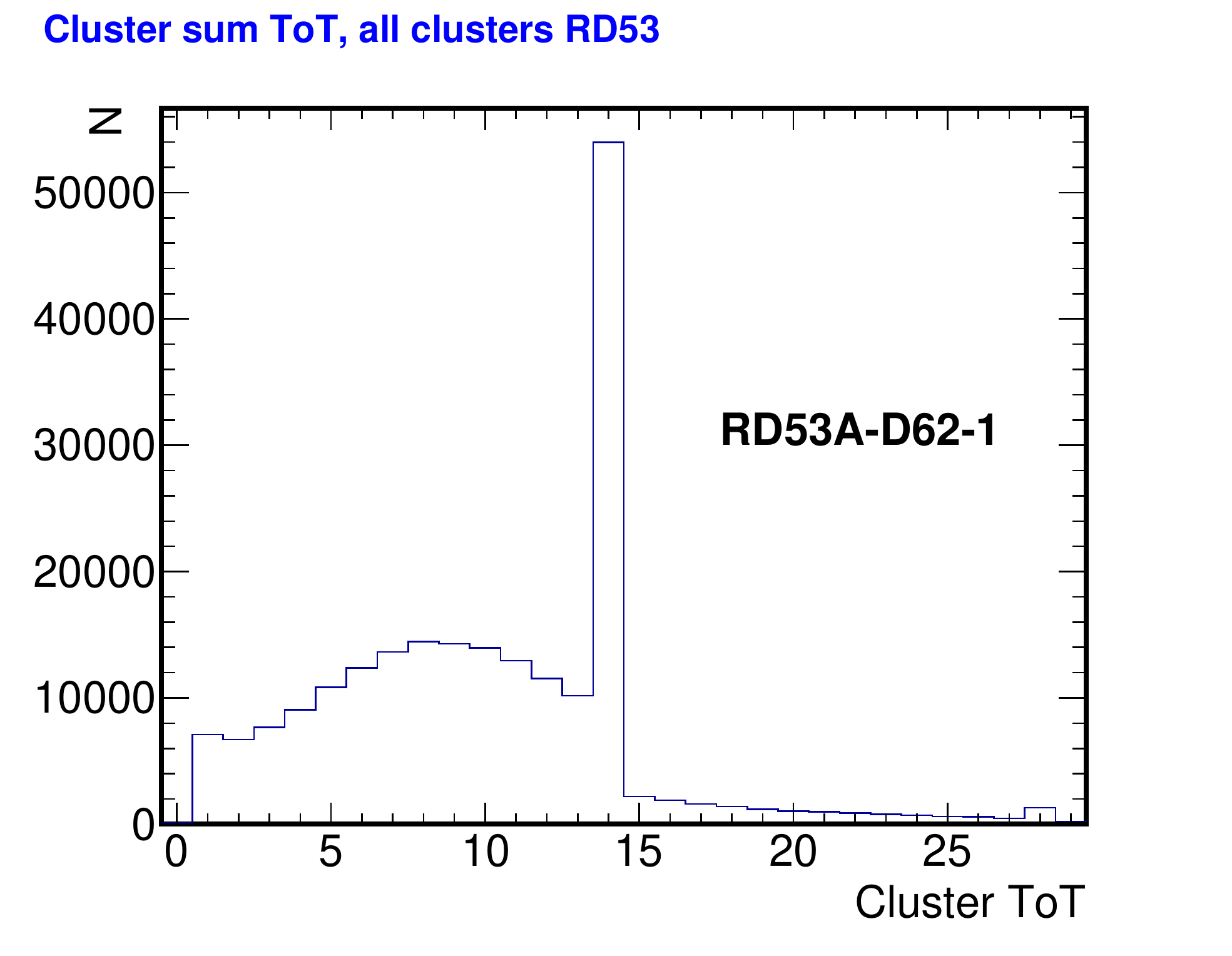}

\end{minipage}
\caption{TOT distributions for clusters of the D61-2 (left) and D62-1 (right) sensors, irradiated to fluences of $5.0 \times 10^{15}$ $n_{eq} cm^{-2}$ and $1.0 \times 10^{16}$ $n_{eq} cm^{-2}$ respectively. The rightmost peaks in the cluster sum ToT plots, in bin 14, are due to saturation at the highest value for the ToT (14) in a single pixel.}
\label{fig: DESY_irrad_both}
\end{figure}

Samples were irradiated to fluences of $5 \times 10^{15}$ $n_{eq} cm^{-2}$ and $1.0 \times 10^{16}$ $n_{eq} cm^{-2}$, and efficiencies were also studied as functions of bias voltage. In this study, the tracks passed through the sensors at normal incidence. Figure \ref{fig: DESY_irrad_both} shows the ToT distribution for the DUTs D61-2 (left) and D62-1 (right), with a threshold tuned to 1 k$e^-$ and 0.5 k$e^-$ respectively. The different shapes of the distributions is a consequence of the difference in thresholds where the distribution for D61-2 in Figure \ref{fig: DESY_irrad_both} have been cropped by a too high threshold setting. The saturated bin (bin 14) in both distributions corresponds to a long tail as expected from the Landau-Vavilov distribution.

\begin{figure}[!htb]
\centering
\begin{minipage}{1\textwidth}
    \centering
    \includegraphics[width=1.0\linewidth]{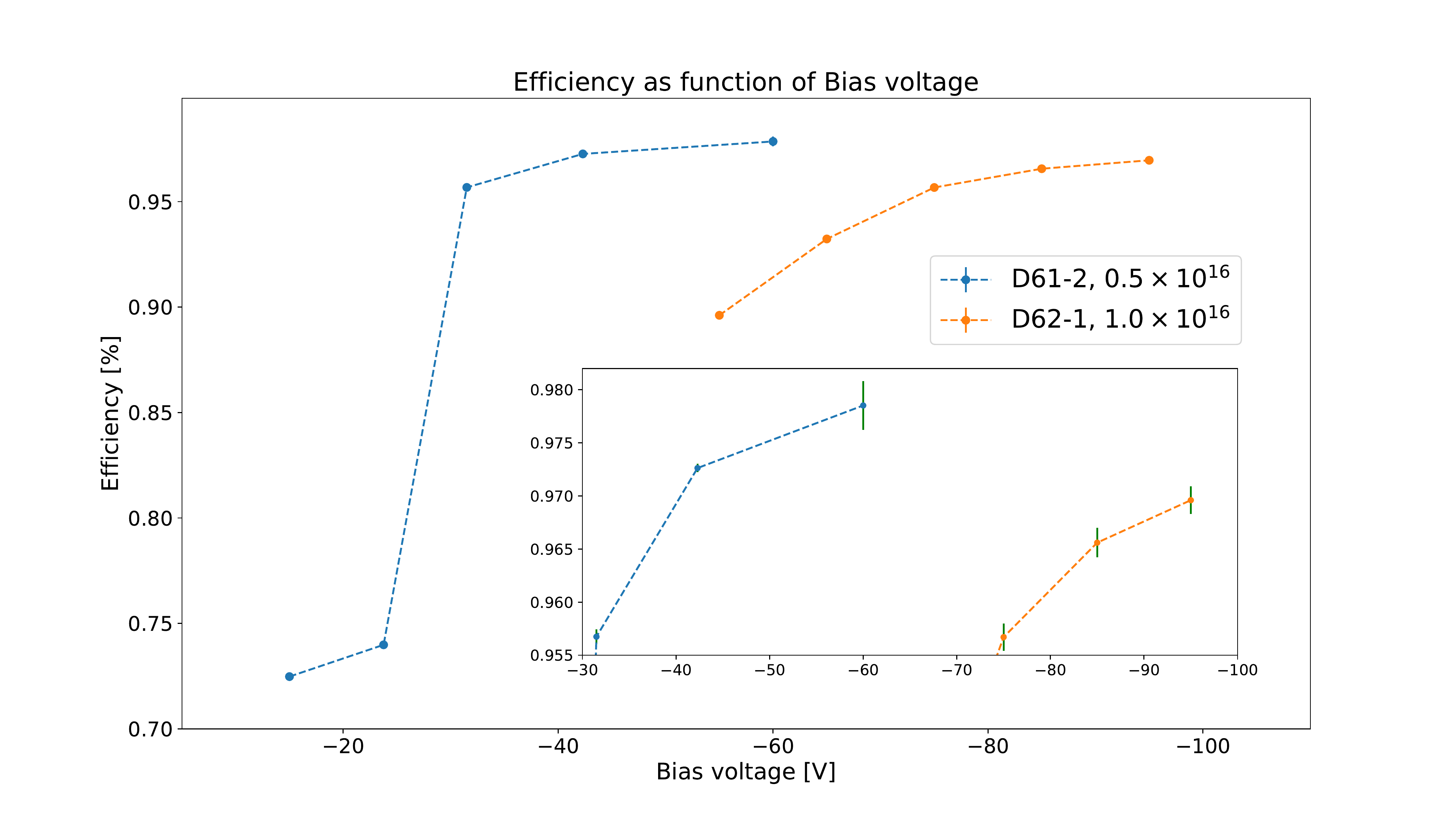}
    \caption{Efficiency as a function of applied bias-voltage for two irradiated sensors. Irradiation to a fluence of $5\times10^{15}n_{eq}/cm^2$ (blue), and $1\times10^{16}n_{eq}/cm^2$ (orange). The inset is a zoomed view of the last three measurements for each sensor to see the uncertainties.}
    \label{fig:D61-2_D62-1}
\end{minipage}%
\end{figure}

Figure \ref{fig:D61-2_D62-1} shows the  efficiency as a function of applied bias voltage. An efficiency above 97\% is reached for a bias of about 40 V for sensor D61-2 and above 96.5\% at about 80V for D62-1. 

%% file: Sections/SummaryConclusions.tex
\section{Summary and Conclusions}
We have described the production of 3D sensors at SINTEF of the 'RD53A' design, and presented results of subsequent tests in pion and electron 
beams for unirradiated sensors as well as for sensors irradiated to fluences of 
$5 \times 10^{15}$ $n_{eq} cm^{-2}$ and $1.0 \times 10^{16}$ $n_{eq} cm^{-2}$.
It is demonstrated that unirradiated sensors of 50 {\textmu}m and 100 {\textmu}m active thickness
can be operated at efficiencies above 98.6\% for tracks at normal incidence
for a bias of 10 V. 

According to the technical design report for the ATLAS ITk pixel detector \cite{PixelTDR}, a minimum hit efficiency of 96\% is required for pixel sensors in the innermost layer of the ITk after an irradiation to the target fluence of $1.0 \times 10^{16}$ $n_{eq} cm^{-2}$ at normal beam incidence and with an applied bias voltage below 150 V. The results presented in this paper show that the tested samples reach an efficiency above 97\% for an irradiation of $0.5 \times 10^{16}$ $n_{eq} cm^{-2}$ with a bias voltage of V$_{bias}$ = 40 V, and above 96.5\% with a bias voltage of V$_{bias}$ = 80 V after an irradation of $1.0 \times 10^{16}$ $n_{eq} cm^{-2}$. 

In conclusion, the performance of the tested sensors are above the minimum requirements of the ITk.

%% file: main.bbl
\begin{thebibliography}{10}

\bibitem{ATLASID}
W.~Lucas on~behalf of~the ATLAS~collaboration.
\newblock \emph{ATLAS inner tracking detectors: Run 1 performance and
  developments for Run 2.}, Contribution to ICHEP37, Nucl.Part.Phys.Proc
  273-275 (2016).

\bibitem{StripsTDR}
ATLAS collaboration.
\newblock \emph{Technical Design Report for the ATLAS Inner Tracker Strip
  Detector}, CERN-LHCC-2017-005 ; ATLAS-TDR-025.

\bibitem{PixelTDR}
ATLAS collaboration.
\newblock \emph{Technical Design Report for the ATLAS Inner Tracker Pixel
  Detector.}, CERN-LHCC-2017-021 ; ATLAS-TDR-030.

\bibitem{Parker97}
C.J.~Kenney S.I.~Parker and J.~Segal.
\newblock \emph{A proposed new architecture for solid-state silicon detectors},
  Nucl. Instrum. Meth. A{\bf 395} (1997) 328.

\bibitem{Sintef09}
T.-E.~Hansen et~al.
\newblock \emph{First fabrication of full 3D-detectors at SINTEF}, J. Inst.
  {\bf 4} (2009) P03010.

\bibitem{IBL}
Y~Takubo on~behalf of~the ATLAS~collaboration.
\newblock \emph{ATLAS IBL operational experience}, Contribution to Vertex 2016
  DOI:10.22323/1.287.0004, ATL-INDET-PROC-2016-012.

\bibitem{FEI4}
M.~Barbero et~al.
\newblock \emph{ FE-I4,the new ATLAS pixel chip for upgraded LHC luminosities},
  Proceedings of the IEEE Nuclear Science Symposium and Medical Imaging
  Conference, October 30 November 3,Knoxville, U.S.A. (2010),
  ATL-UPGRADE-SLIDE-2009-319.

\bibitem{Dorholt18}
O.~Dorholt et~al.
\newblock \emph{Beam tests of silicon pixel 3D-sensors developed at SINTEF}, J.
  Inst. {\bf 13} (2018) P08020.

\bibitem{RD53A}
A.~Dimitrievska~A. Stiller and the RD53~collaboration.
\newblock \emph{RD53A: A large-scale prototype chip for the phase II upgrade in
  the serially powered HL-LHC pixel detectors}, Nucl. Inst. Meth. (Proc.
  Suppl.) A958 (2020), 162091.

\bibitem{EUDET}
EUDET collaboration D. Haas~et al.
\newblock \emph{ A Pixelated Telescope for the E.U:Detector R\&D}, Proceedings
  of the International Linear Collider Workshop LCWS 2007, and ILC 2007,
  DESY-PROC-2008-03.

\bibitem{MIMOSA}
J.Baudot et~al.
\newblock \emph{First test results Of MIMOSA-26, a fast CMOS sensor with
  integrated zero suppression and digitized output.}, NSS Conference Record
  (2009) DOI: 10.1109/NSSMIC.2009.5402399.

\end{thebibliography}
